%% file: main.tex
\documentclass[10pt,twoside]{elsart}
\usepackage{epsfig}
\usepackage{amssymb,amstext,amsmath}
\journal{N.I.M.~A}
\psfull
\begin{document}
\input title 
\pagebreak
\tableofcontents
\pagebreak
\input intro

\input minos

\input pmt
\input dark

\input scan

\newpage
\newpage
\newpage
\input smry 
\pagebreak

\input bibl\end{document}

%% file: title.tex
\hfill
\begin{minipage}{0.35\textwidth}
{NuMI-PUB-SCINT-820\\ONUP-2002-??\\[3ex]}
\end{minipage}

\begin{frontmatter}

\title{Testing of Hamamatsu R5900-00-M64 Multi-Pixel PMTs for MINOS 
%\\[2ex]
%Draft 3.0
}

\author[Oxford]{M.~A.~Barker},
\author[Oxford]{A.~De~Santo},
\author[UCL]{P.~Dervan},
\author[Austin]{K.~Lang},
\author[Oxford]{P.~S.~Miyagawa},
\author[UCL]{R.~Saakyan},
\author[UCL]{C.~Smith},
\author[CalTech]{D.~Michael},
\author[UCL]{J.~Thomas},
\author[Oxford]{A.~Weber\thanksref{corr}}

\thanks[corr]{Corresponding Author: A.Weber@physics.ox.ac.uk}

\address[Oxford]{University of Oxford, NAPL, Keble Road, Oxford OX1 3RH,
United Kingdom}
\address[UCL]{University College London, Gower Street, London WC1E 6BT,
United Kingdom}
\address[Austin]{University of Texas at Austin, Austin, Texas
78712-1081, USA}
\address[CalTech]{California Institute of Technology, Charles
C.~Lauritsen Laboratory, Pasadena, CA 91127, USA }

\begin{abstract}
The MINOS long baseline neutrino oscillation experiment is comprised of
three calori\-meters, a Near Detector at FNAL, a Far Detector at the
Soudan mine in northern Minnesota, and a Calibration Detector in a test
beam at CERN. 
The light
produced in the interactions of neutrinos 
in the Near Detector at FNAL will be routed
by optical fibres onto the pixels of Hamamatsu R5900-00-M64 multi-anode
photomultiplier tubes.  This article reports the measurements performed
on 15 of these tubes to evaluate them for their use in MINOS.
\end{abstract}

\end{frontmatter}

%% file: intro.tex
\section{Introduction}
\label{sec:intro}
The MINOS (Main Injector Neutrino Oscillation Search) Experiment has
been designed to identify neutrino oscillations\cite{nuosc-expts} and
measure their parameters in the controlled environment of the NuMI
neutrino beam at Fermi National Accelerator Laboratory (FNAL). The beam of
predominantly muon neutrinos will pass through the Near Detector (ND)
a calorimeter composed of planes of plastic scintillator and steel,
at FNAL where the original flavour composition of the beam will be
measured. The Far Detector (FD), a 5.4KT calorimeter of similar
composition, situated in
the Soudan mine in Northern Minnesota, will measure the flavour
composition of the same beam $735$~km further away.

Both detectors will use extruded scintillator strips, read out via
wavelength shifting (WLS) fibres  glued into a central groove, to detect
energy depositions of particles traversing the detector.  The light from
the WLS fibres is transmitted by clear acrylic fibres to 16-pixel
Hamamatsu R5900-00-M16\cite{m16} photomultiplier tubes (PMTs) in the FD
and to 64-pixel Hamamatsu R5900-00-M64 (M64) PMTs in the ND.  Two different
types of PMTs are required in order to take economical advantage of 
optical summing, possible in the low rate
environment of the FD
but not in the high-rate environment of the ND.

The results from testing fifteen M64s are reported in this paper.
%have been tested to gain experience with these
%tubes and to understand their production variations. 
%A procurement
%agreement with Hamamatsu was based on the data presented in this
%article. 
%The results of the testing of all 230 tubes for the MINOS
%experiment will be published in a separate paper.

The article is organised as follows: Section \ref{sec:minos} describes
the requirements on the tubes from the MINOS experiment while Section
\ref{sec:m64} gives technical details about the M64. 
Results from dark noise
measurements are given in Section \ref{sec:dark-results},
while Section\ref{sec:scan-results} describes the details of the
measurements of different PMT properties. 

%, while after-pulsing is discussed in Section
%\ref{sec:afterp}.

%% file: minos.tex
\section{Required performance of PMTs for MINOS}
\label{sec:minos}

The main requirements on the PMT characteristics are the following:
\begin{itemize}
\item[-]
good single photoelectron (p.e.) sensitivity;
\item[-]
linearity up to 200 p.e.;
\item[-]
single-channel gain above $3\times 10^5$ and up to $10^6$;
\item[-]
small inter-pixel cross-talk;
\item[-]
uniformity of response that does not significantly change 
in the presence of a magnetic field of up to $5$~gauss;
\item[-]
tube dark count rate of less than $3$~kHz above a $1/3$ p.e. threshold.
\end{itemize}

In order to achieve the proposed physics goals, MINOS has to detect
minimum ionising particles (MIPs) and has to measure electromagnetic
showers. For a MIP, the energy released in the
scintillator corresponds to about $2-8$ p.e.\ per
strip traversed. On the other hand, electromagnetic showers can produce
light signals as large as 200 p.e. It is therefore of the utmost importance
that the photodetectors used have a wide dynamic range: they must be
able to measure single p.e.\ signals precisely while still behaving as
linear devices at light levels of up to 200 p.e.

The digitisation of the PMT signals is carried out roughly 1~m away from
the tubes. Therefore, to avoid any effect due to electromagnetic pick-up
noise, the PMT gain should be greater than $3\times 10^5$.

Due to the high beam intensity at the near detector, up to $200$
neutrino events may be
present within the same $11~\mu$s spill shortly after the primary
protons hit the neutrino production target. To minimise complications in
the event reconstruction algorithms, cross-talk to all neighbouring
pixels of the same PMT should be below 10\%.

The PMTs will be located close to the magnetised steel planes, where a
fringe field of about $5$~gauss is expected. The PMT will be installed
inside a steel housing, which will act as a shield and should reduce the
field by at least $80\%$.  However, it is important to verify that the
tube response is not significantly affected by the remaining field and
to have a safety margin.

All hits are continuously digitised by the front-end electronics and
sent via readout processors to a ``trigger farm'', which decides,
depending on the event pattern, whether or not to save the data for further
analysis. As the processing power and the bandwidth of the transfer 
are limited, it is important that the performance of MINOS is not
affected by the presence of noise.  The dark count rate of a
single PMT channel should therefore be limited to about $50$~Hz. This is
mainly due to bandwidth limitations of the DAQ system.  The digitisation
of all anode channels is started by the presence of a sizeable signal on
the common last dynode. It is therefore important to study the noise
properties of the anodes as well as of the last dynode.

%% file: pmt.tex
\section{Technical characteristics of M64 PMTs}
\label{sec:m64}

Before describing in detail the procedure to evaluate the performance 
of the PMTs, it is useful to summarise the technical characteristics of
the average M64 PMT as provided by Hamamatsu
Photonics\cite{hamamatsu-address}.

The Hamamatsu R5900-00-M64 is a $12$-dynode multi-anode PMT, with a
transverse section of $1\times 1$~inches.
% (see fig.~\ref{fig:m64pic}).
The $64$ pixels (2~mm~$\times$~2~mm) are arranged on an $8\times 8$
square grid. These pixels are separated from each other by a distance
of less than 0.3~mm, but there is a region of reduced collection
efficiency extending to 0.3~mm.

Focusing electrodes run over the pixels
which guide the photoelectrons into the dynode
structure. Each pixel has two identical dynode chains associated with it
which terminate in a single anode pad.

The cathode and the housing of the tube are at negative potential and a
tapered bleeder circuit is used to apply voltage to the
dynodes. The divider ratio adopted for the
present measurements is the one recommended by Hamamatsu for optimal
performance ($3:2:2:1:1:1:1:1:1:1:1:2:5$). Decoupling capacitors at the
last stages help to keep the potential differences constant between the
last dynodes and the anode in the presence of large instantaneous currents.
The main parameters for the average PMT, as provided by
Hamamatsu, are summarised in Table~\ref{tab:hamamatsu}.

\begin{figure}
\centerline{
\includegraphics[width=0.8\textwidth]{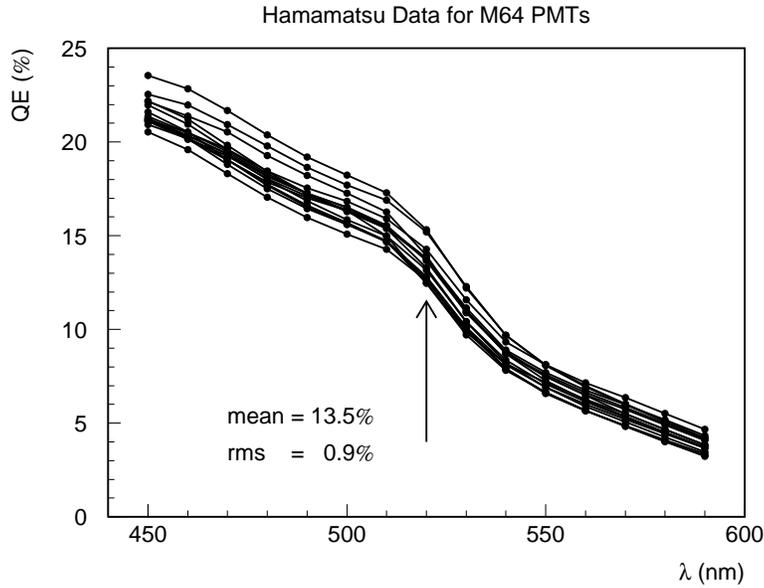}
}
\caption{QE vs. incident light wavelength for the 15 M64 PMTs
as measured by Hamamatsu. 
The arrow marks the position of $\lambda=520$~nm.}
\label{fig:qe_vs_lambda}
\end{figure}

The only data available concerning the quantum efficiency (QE) of the
PMTs are those given in the Hamamatsu data sheets.  Fig.~\ref{fig:qe_vs_lambda}
shows QE as a function of the incident light wavelength $(\lambda)$ for
$15$ tubes.  The mean value of QE at $\lambda=520$~nm,
the wavelength of interest for MINOS, is 13.5\% with a spread of 0.9\%.

\begin{figure}
\centerline{
\includegraphics[width=0.8\textwidth]{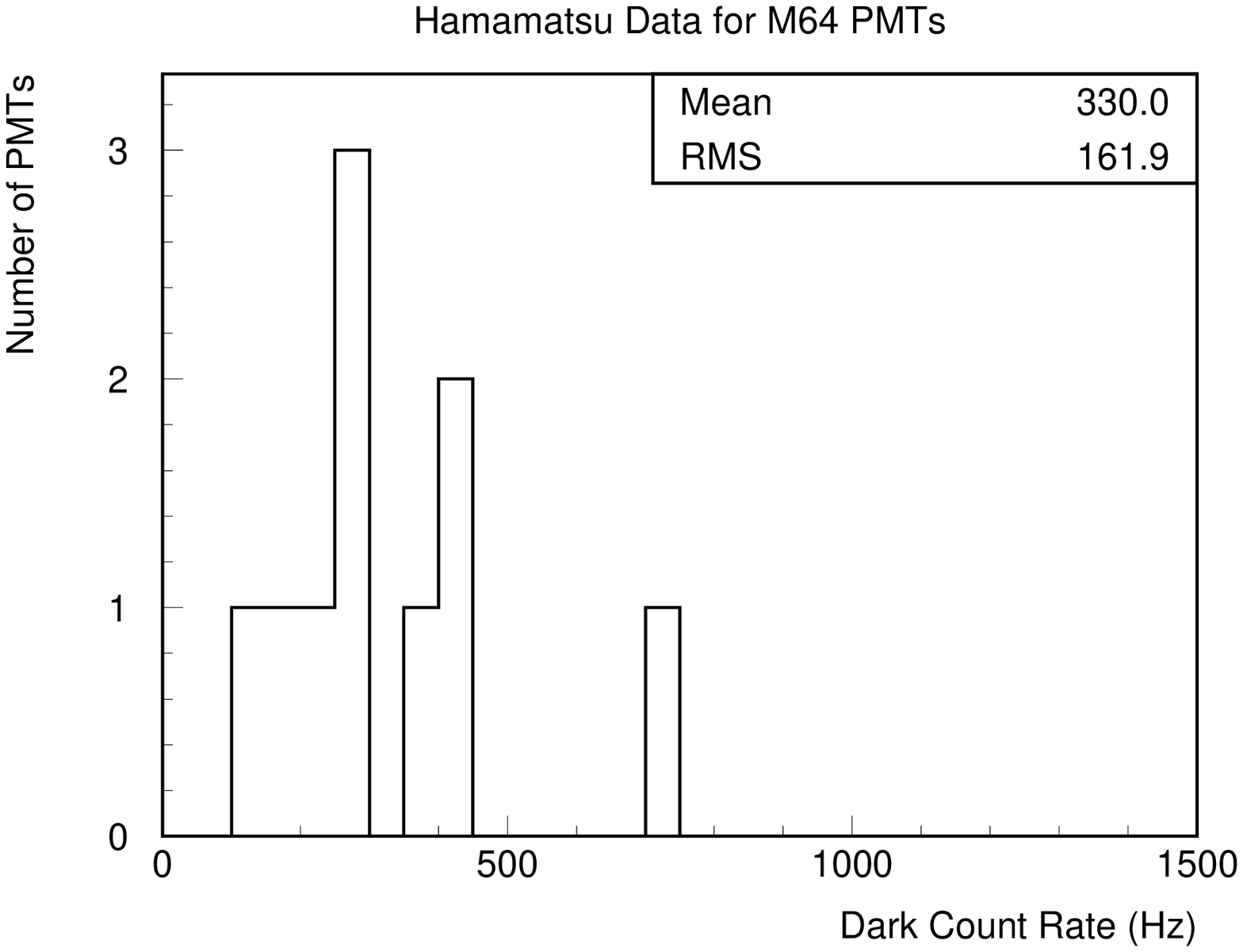}
}
\caption{Dark count rate of 10 M64 phototubes, as measured
by Hamamatsu $($at $900$~V and after $30$~minutes in the dark$)$.}
\label{fig:ham_dark_noise}
\end{figure}

For ten of the tubes, Hamamatsu measured the anode dark count rates at $900$~V
and after $30$ minutes in the dark.  The distribution of measured rates is
shown in Fig.~\ref{fig:ham_dark_noise}. Integrating over all the
channels, the dark count rate is approximmately
$330$~Hz. This is well below the maximum rate acceptable for MINOS.

%\begin{figure}
%\centerline{
%\includegraphics[width=\textwidth]{plots/m64-drawing.eps}
%}
%\caption{Mechanical layout of a Hamamatsu R5900-00-M64 PMT.} 
%\label{fig:m64pic}
%\end{figure}

\begin{table}
\begin{center}
\begin{tabular} {||l|c||}
\hline
\hline
				&				\\
Parameter 			& Description/Value 		\\
				&				\\
\hline \hline
Photocathode material		& bi-alkali 			\\ 
Spectral response		& 300 to 650~nm 		\\ 
%Quantum efficiency (QE) at 520 nm& $\geq$ 13 \% 		\\
%Quantum efficiency (QE) at 520 nm& $\geq$ 12.5 \% 		\\
Quantum efficiency (QE) at 520 nm& $\geq$ 12.0 \% 		\\
Wavelength of maximum QE        & 420~nm 			\\
Window material  		& borosilicate glass		\\
%Index of refraction of window	& 1.492 @ 486 nm		\\
%(reference data)		& 1.486 @ 589 nm		\\
Window thickness		& 0.8~mm 			\\
Material of metal casing	& KOVAR				\\ 
Weight				& $(27.9\pm0.1)$~g 		\\
Dynode type			& metal channel structure 	\\
Number of stages		& 12 			 	\\ 
Anode				& array of 8 $\times$ 8 independent pixels \\
Anode dark current per channel  & $\leq$ 0.2 nA 		\\
Pixel size                      & 2 mm $\times$ 2 mm square 	\\
Maximum high voltage		& 1000 V 			\\
Gain at 800 V (typical)		& $^{(1)}$$\sim$3$\times$10$^5$	\\ 
Anode pulse rise time		& 1.5 ns 			\\
Transit time spread per channel (FWHM)& 0.3 ns  		\\
Pulse linearity per channel ($\pm$ 5 \%)& $^{(1)}$ 0.6 mA \\
Uniformity between each anode	& 1:3 				\\
\hline
\hline
\end {tabular}
\end {center}

\caption{\label{tab:hamamatsu} Basic nominal characteristics of the 
R5900-00-M64 64-channel 
photomultiplier produced by 
Hamamatsu Photonics~\cite{M64_spec_sheet}.
$^{(1)}$ The voltage divider ratio is assumed to be
3-2-2-1-1-1-1-1-1-1-1-2-5).}
\end {table}

%% file: dark.tex
\section{Dark Noise}
\label{sec:dark-results}

The light signal released by a minimum ionising particle in MINOS 
can be as low as two photoelectrons.
It is therefore extremely important to study the PMT dark noise 
at very low threshold levels. Since the last dynode signal will
be used in MINOS to trigger the anode readout, the 
dark count rate has to be measured for dynode thresholds
corresponding to anode signals of approxiately  $1/3$ photoelectron.

\subsection{Set-up description}

\begin{figure}
\begin{center}
\epsfig{file=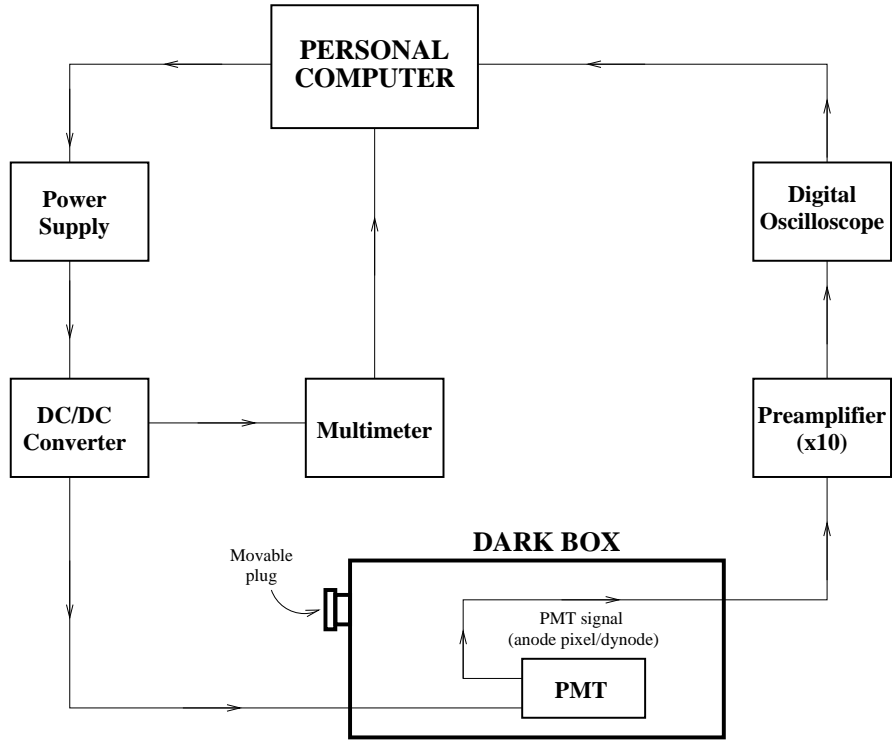,width=0.9\linewidth}
\caption{Schematic of the experimental dark count setup}
\label{fig:setup-dynode}
\end{center}
\end{figure}

A block scheme of the experimental set-up used for these measurements 
\cite{alfons} is shown in
Fig.~\ref{fig:setup-dynode}.  The PMT under test is connected to its
base and placed into a dark box. A removable plug is used to allow light
into the otherwise light-tight box.  The high voltage is supplied to 
the PMT by a Hewlett-Packard DC
power supply (HP-E3631A) through a DC/DC converter circuit and
monitored by a digital multi-meter (HP-34401A). Both are connected to a
PC ($500$~MHz, Pentium III processor) through a GPIB interface. 

The signals from the last dynode and from the anode pixel under study
are sent via a $50~\Omega$ coaxial cable and an amplification stage
(Phillips 779, nominal gain $10.0\pm0.2$) to a four-channel
Tektronix digital oscilloscope (TDS~540A).  All the remaining anode
channels are terminated through $50~\Omega$ to prevent charge
build-up. The oscilloscope bandwidth is limited to 100~MHz to reduce noise.

The PC communicates with the oscilloscope via a LabVIEW based program.
Once the triggering conditions have been set and the channel whose
data will be acquired has been selected, the program reads in 
the waveform of the PMT pulse event by event and measures:
\begin{itemize}
 \item[-]
the baseline level, to account for any DC offset by the amplifier;
 \item [-]
the signal amplitude.
\end{itemize}
The oscilloscope acquires 910 pulses at a rate in excess of 10 kHz before
transferring the data to the PC. The oscilloscope is inhibited during
the transfer which takes about one second.
 
To obtain an absolute measurement of the gain, although not
crucial for the determination of the dark count rate, 
a voltage-to-charge conversion factor of  $(0.81\pm 0.16)\times
10^{-11}~\mbox{C}/\mbox{V}$ has been measured by
assuming shape universality for the PMT pulses.

\subsection{Anode dark noise}

\begin{figure}
\centerline{
\includegraphics[width=\textwidth]{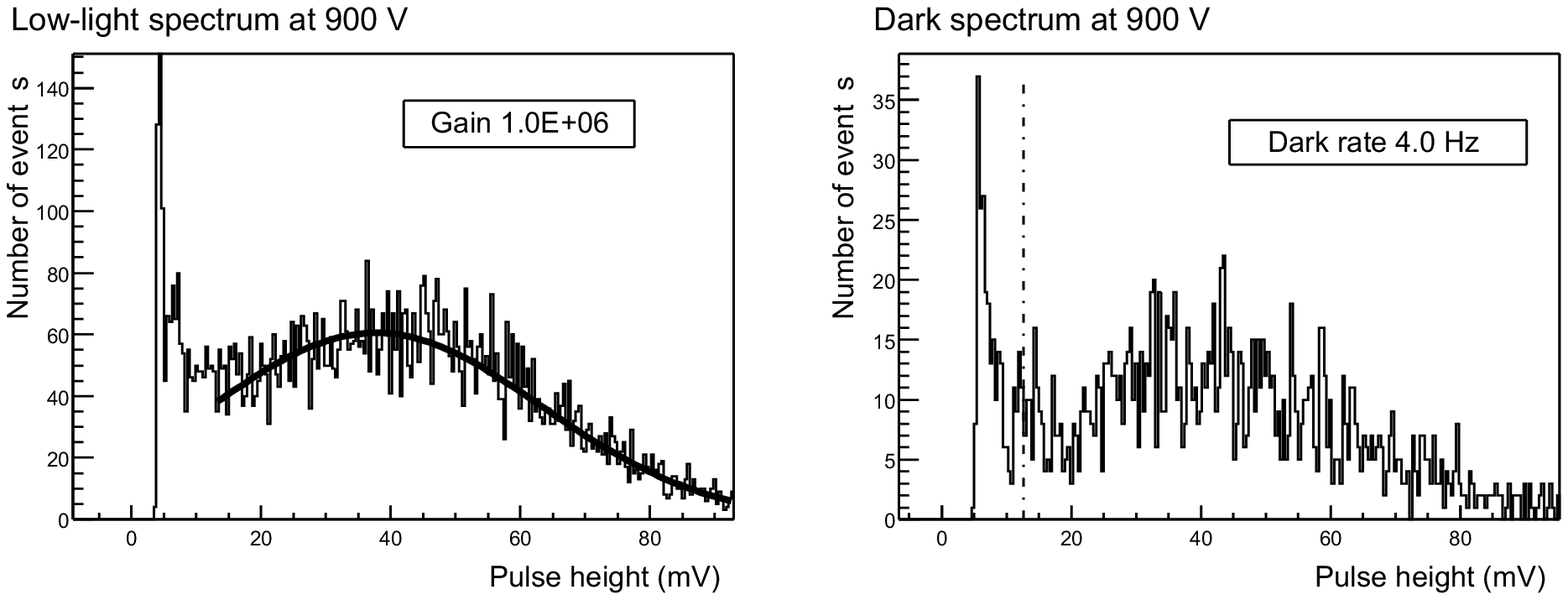}
}
\centerline{
\includegraphics[width=\textwidth]{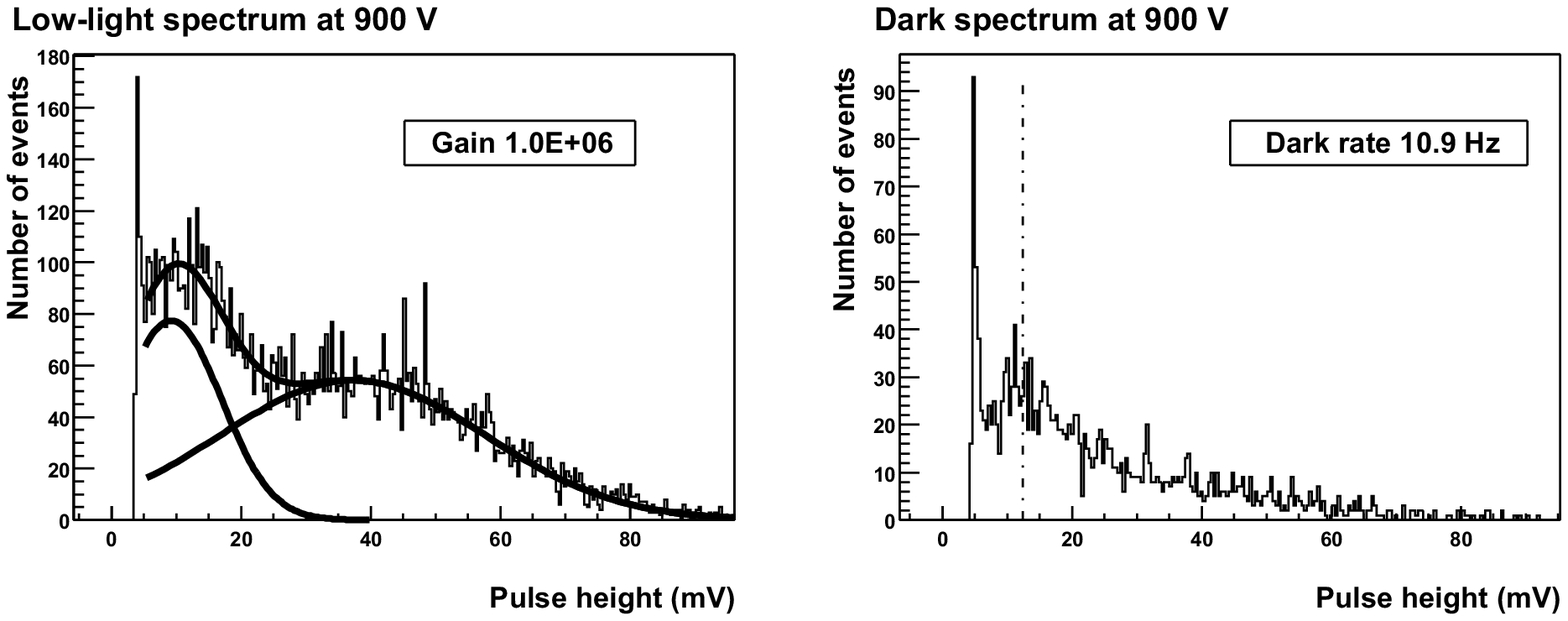}}
\caption{Low-light and dark spectra for an inner (top) and an outer
(bottom) pixel.  
The dark count rate was determined using an offline threshold set to
$1/3$ of a photoelectron, as shown by the dotted line. Note the
appearance of a peak between the noise and single-photoelectron peak for the
outer pixel.}
\label{fig:dark}
\end{figure}

Each of the tubes under test was kept in the dark and 
under high voltage ($900$~V) for at
least six hours before performing the actual measurement. 
Low-light spectra and dark
spectra were taken at high voltages between $900$ and $1000$~V 
for all pixels in one quadrant of the tube. The oscilloscope
trigger threshold was set just above the anode pedestal peak which is at
zero, as evidenced in Fig.~\ref{fig:dark} which shows the falling edge of
the pedestal at very low pulse heights. The dark count rate was
determined using an offline threshold of 1/3 of a photoelectron.

For all of the measured tubes, the low-light spectra of the pixels
belonging to the inner $6 \times 6$ matrix show a distinct
single-photoelectron peak,
well separated from the electronic noise. These pixels typically have
relatively low dark count rates (a few Hz). Typical low-light and dark
spectra of such pixels are shown in Fig.~\ref{fig:dark}.  Pixels along
the edges of the PMT generally have higher noise rates, although in most
cases still within acceptable limits. Of greater concern is the shape of
their low-light spectra, since quite frequently the single-photoelectron
peak is not
very well distinguishable from the dynode noise contribution. Moreover a
second peak, distinct from the noise but at pulse heights below the
single-photoelectron peak, is often present (see Fig.~\ref{fig:dark}).  
This is due to photons passing through the photocathode and hitting
the first dynode where they are absorbed and a single photoelectron is
emitted. This results in a peak in the distribution
corresponding to an amplification of one dynode stage less than the
main one photoeletron peak. The different
behaviour between the inner and the outer pixels is expected to be due to
different focusing conditions as well as to the difficulty in keeping the
fabrication process uniform across the entire photocathode area.

\begin{figure}
\centerline{
\includegraphics[width=0.7\textwidth]{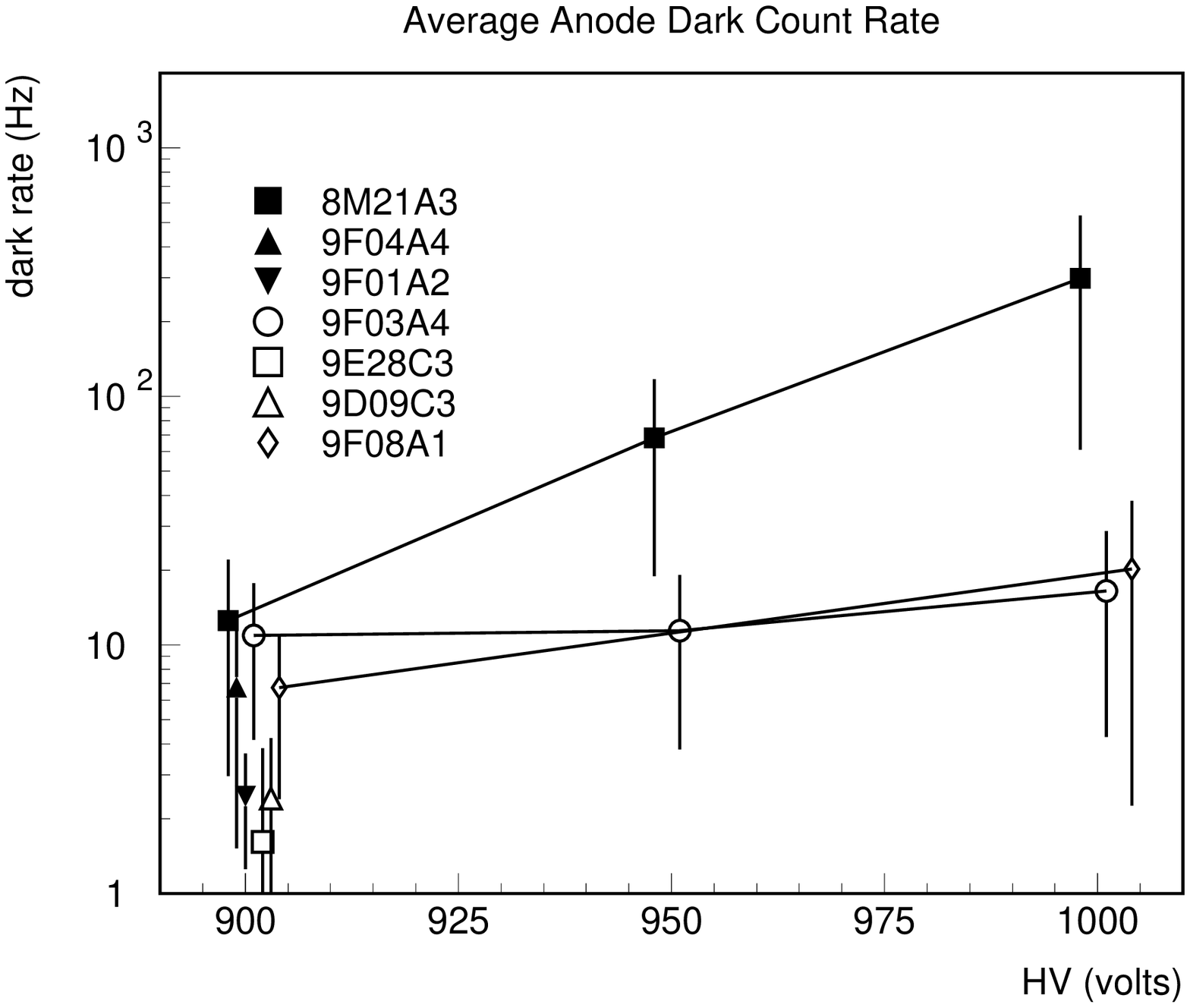}}
\caption{Voltage dependence of the dark count rate.  Single pixel averages
with RMS distribution are shown for each tube.  While voltages of 900~V,
950~V and 1000~V were always used, the data points have been spread out
in order to make them easier to distinguish.}
\label{fig:dark-volt}
\end{figure}

The dependence of the dark noise on the PMT high voltage has also been
studied. This is shown in Fig.~\ref{fig:dark-volt}. The dark count rate
increases at higher values of high voltage. The single anode dark count rate at
$900$~V is measured to be $5.2$~Hz, with a spread of $3.0$~Hz.

\begin{figure}[bt]
\centerline{
\includegraphics[width=0.7\textwidth]{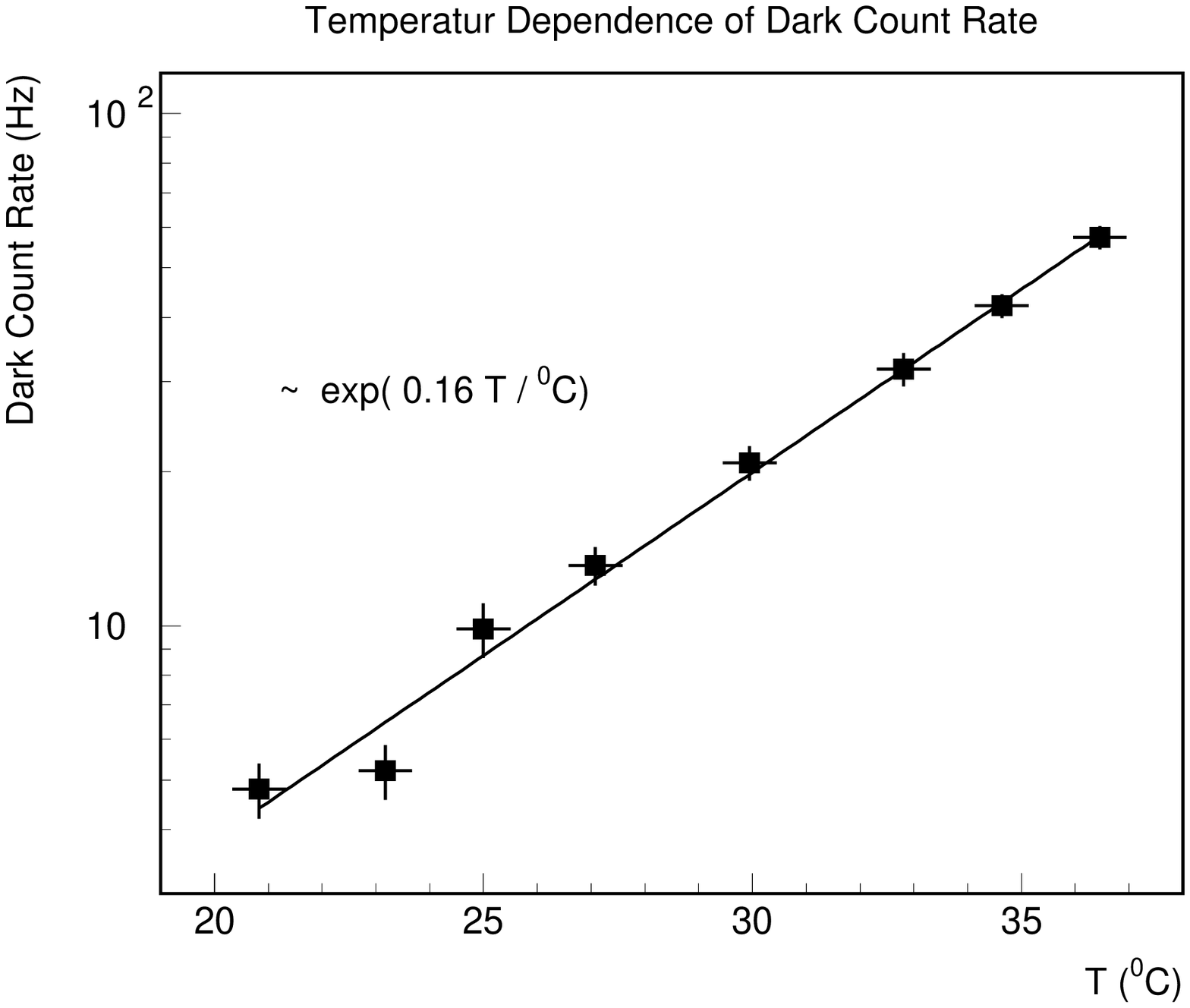}}
\caption{Temperature dependence of dark count rate.}
\label{fig:dark-temp}
\end{figure}

The temperature dependence of the dark count rate has been measured for
one PMT. For both inner and outer pixels the gain does not vary
substantially with temperature, while, as expected for bi-alkali
photocathodes, the dark count rate increases exponentially with
temperature (($16 \pm 3) \: \%$~per\ $^\circ$C; see
Fig.~\ref{fig:dark-temp}).

\subsection{Dynode dark noise}

The last dynode is common to all anodes, and so its single-photoelectron
spectrum can
be viewed as the sum of $N_{pixel}$ charge distributions, each of them
corresponding to the single-photoelectron signal on one of 
the $N_{pixel}$ pixels. As
the gain is different for each channel, in general no clear peak 
is visible when the spectrum is acquired by triggering on the
dynode signal itself.

\begin{figure}
\begin{center}
\epsfig{file=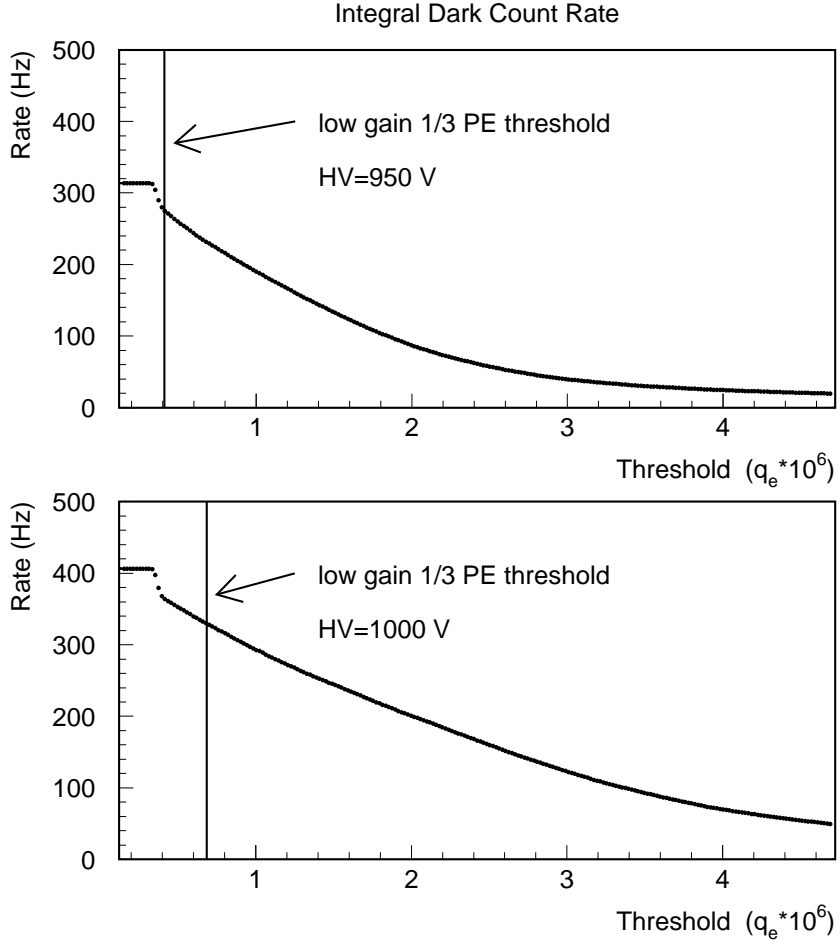,width=0.9\linewidth}
\caption{Integral dynode dark count rate for the tested M64
as a function of the trigger threshold, at $\mbox{HV}=950~\mbox{V}$ (top) 
 and $\mbox{HV}=1000~\mbox{V}$ (bottom). The vertical lines indicate the
 threshold corresponding to $1/3$~p.e.\ of the lowest gain pixel.}
\label{fig:m64_int}
\end{center}
\end{figure}

It is necessary to determine the noise rate at the dynode 
caused by signals above $1/3$~p.e. from the lowest gain pixel.
To do this, the PMT was illuminated at a very low level.
Initially, the spectrum from the lowest gain pixel was digitized and
confirmed to be mostly one p.e. signals. Then,
the pulses from this pixel were used to trigger the oscilloscope 
while the dynode signal was digitized. This enabled us to measure the
contribution to the dynode spectrum from this channel and to determine
the dynode threshold level which corresponded to $1/3$~p.e. on the
lowest gain pixel as shown in Figure \ref{fig:m64_int}.

After light-sealing the dark box again, the PMT was kept under high
voltage and in the dark for more than 24 hours.  A dark spectrum of the
last dynode was taken, triggering on the dynode signal itself. In order
to reach the necessary sensitivity to the dark count rate, the trigger
threshold had to be tuned somewhat below the $1/3$~p.e.\ level
ascertained in the previous measurement, but high
enough not to be dominated by electronic noise.  Knowing the acquisition
time, the integral dynode dark count rate as a function of the trigger
threshold can be computed from the pulse height distribution.
Knowing the relative positions of the anode and of the last dynode 1~p.e.\ 
peaks, the gain of the last amplification stage can be
computed from simple charge conservation considerations. At $950$~V, the
gain of the last dynode is $\approx3.5$.

The integral dynode dark count rate as a function of the trigger
threshold, which was obtained for both settings of high voltage, is shown in
Fig.~\ref{fig:m64_int} for one of the PMTs. The count rate for a
threshold of $1/3$~p.e.\ is equal to about $275~\mbox{Hz}$ at
$950~\mbox{V}$ and to about $330~\mbox{Hz}$ at $1000~\mbox{V}$; as
expected, the higher the gain, the higher the dark noise.

\begin{figure}
\begin{center}
\epsfig{file=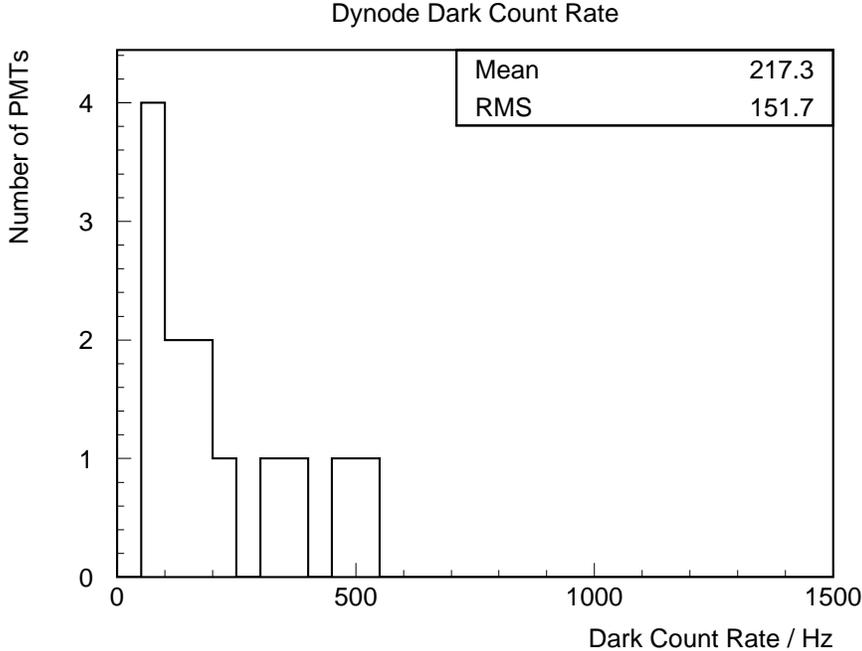,width=0.9\linewidth}
\caption{Dynode dark count rate at with a $1/3$~p.e.\ threshold.}
\label{fig:dynode_dark_count}
\end{center}
\end{figure}

Following the procedure described above, the last dynode dark count
rate at $1/3$~p.e. has been measured for all the available PMTs 
at $950$~V and after $\gtrsim 24$~hrs in the dark. The
resulting distribution for the 15 tubes is shown in 
Fig.~\ref{fig:dynode_dark_count}. The mean value is about $220$~Hz,
which confirms that the tubes under test are extremely quiet.

%% file: scan.tex
\section{Results from PMT scanning}
\label{sec:scan-results}

Most of the properties of the PMTs have been studied by measuring the
phototube response to a point-like light source, which has been moved 
across the photocathode surface. This procedure is referred to as a
``PMT scan''.

\subsection{Set-up}

\begin{figure}
\centerline{
\includegraphics[width=\textwidth]{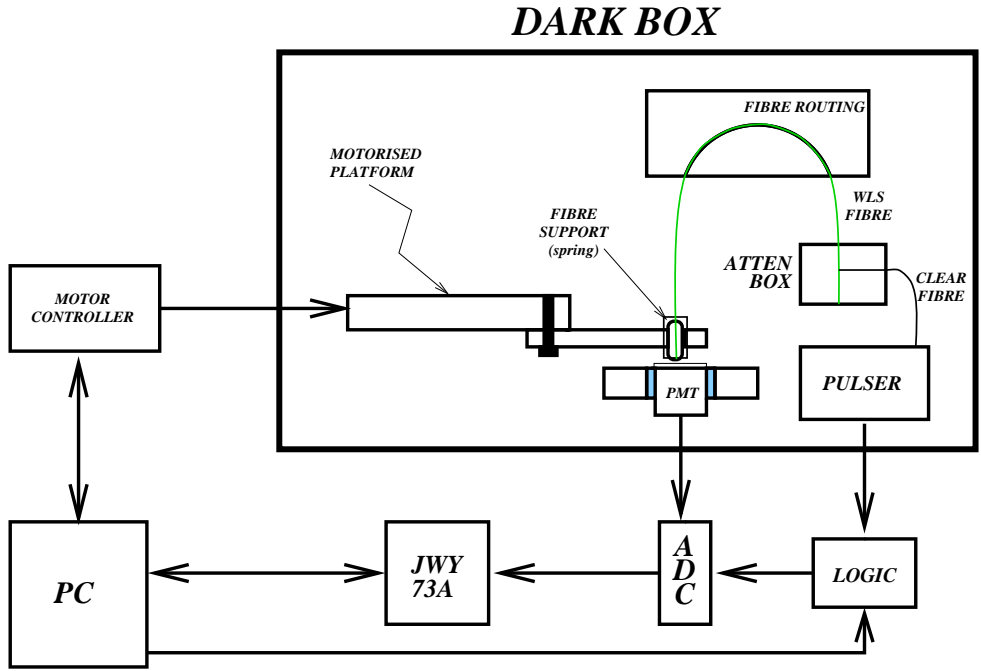}
}
\caption{Block diagram of the set-up used for the PMT scanning.}
\label{fig:scan_setup}
\end{figure}

Two different set-ups have been independently developed, and most
phototubes have been tested in both setups.  While in one set-up the
PMT was moved to scan the light source across the photocathode, the
other assembly used a moving plastic fibre.  The results achieved with
both setups are consistent. We therefore only describe the latter 
apparatus in detail. A schematic view of the setup with the movable fibre is
shown in Fig.~\ref{fig:scan_setup}~\cite{alfons}.

The PMT under test is mounted into the scanning apparatus
with the help of a tightly fitted plastic collar, which in turn is
mounted onto an aluminium PMT holder.
In order to prevent bad connections between the PMT and the base
sockets, the base is screwed to the PMT holder from the back.

Light output from a programmably pulsed blue LED at a repetition rate of
16~kHz is used to illuminate the PMT. The blue light is carried by a
clear acrylic fibre to an attenuation box. In this box the clear fibre
is mounted at a fixed distance (90 degree angle) to a 1 m long WLS
fibre\cite{kuraray} in a similar LED illumination scheme as
used in the MINOS scintillator modules. The WLS fibre 
carries the light to the PMT surface.
For the measurements described here, the level was set to be between a
few and 200 photoelectrons. To prevent possible transmission changes in
the WLS fibre while moving it across the PMT cathode, a routing device
was used to constrain the fibre to slide in an
arc-shaped groove ($25$~cm diameter, 0.5~cm width).  
The fibre was connected to a
non-magnetic armature that is in turn screwed to a motorised platform. 
This arrangement moves
the fibre across the photocathode. The fibre itself is fitted into a
black plastic cylinder which is gently pressed against the glass window
by a spring.

The high voltage is supplied to the PMT as in the dark noise set up.
The motorised platform is powered by a SmartDrive SA46 base unit,
containing two D28 drives. The drives are controlled by a Trio Motion
Technology Euro205 Motion Coordinator linked to the PC via a serial
interface.

The 64 PMT anodes are connected to LeCroy 2249A ADC modules ($10$ bits,
$0.25$~pC/count sensitivity); these are read out by a Jorway 73A CAMAC
crate-controller coupled to the PC via a SCSI connection. Trigger
signals from the LED pulser, coinciding with the light pulses, are used
to gate the ADCs.  The whole set-up is controlled by a LabVIEW 5.1
program and events are taken at a rate of $1.5-2$~kHz, i.e., not every
light pulse is digitised.

The standard program of measurements for a PMT consists of the
following steps:
\begin{enumerate}
  \item an alignment run, to get the PMT centre coordinates with respect
    to the scanning table home position;
  \item $8\times 8$ point scans (WLS fibre at the centre of the pixels) 
    at different high voltages ($U=$ 850, 900, 950~V) and
    light levels (from a few to 100 photoelectrons);
  \item determination of the lowest gain pixel and measurement of
      the dynode dark count rate at $1/3$ p.e. 
      (see Section~\ref{sec:dark-results}). 
\end{enumerate}
At the beginning of each scan a pedestal run is acquired, to determine 
the zero offset for each ADC channel. 
For each fibre position $10,000$ events were recorded, and
the ADC information for all of the $64$ channels was stored as 
one-dimensional histograms.

\subsection{Tools and Techniques}
\subsubsection{Alignment}

Every time a tube was installed in the PMT holder, an alignment
run was performed before running the actual scan in order
to determine the
position of the PMT centre with respect to the home position of the
scanning table. The home position is known to within $10~\mu$m.

\begin{figure}
\centerline{
\includegraphics[width=\textwidth]{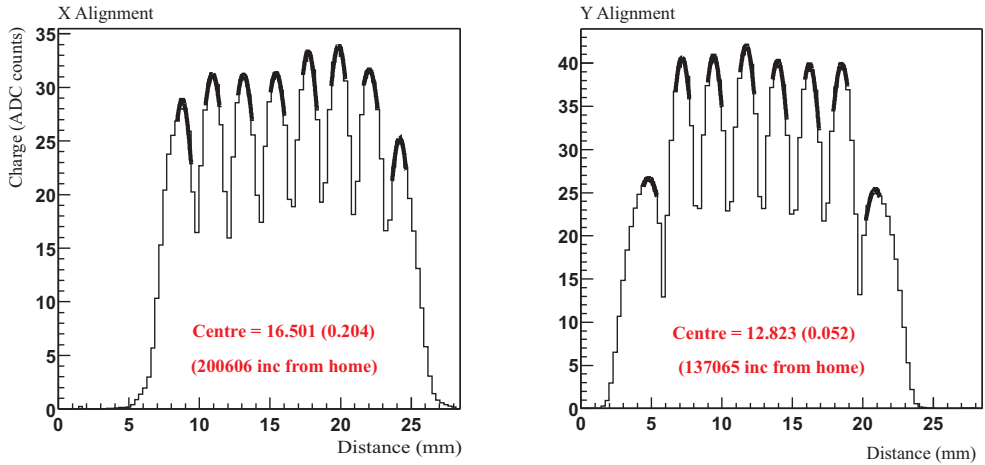}
}
\caption{Typical profiles obtained from an alignment run $($PMT
GA0145 in this case$)$.}
\label{fig:align_profile}
\end{figure}

The alignment run consists of two scans, one along a row of the PMT
$(x$-axis$)$ and the other along a column $(y$-axis$)$, with a step size
of $\approx 250~\mu$m $(100$ points per scan$)$ and with statistics of
$10,000$ events per point.  For each point, the pulse height in the
pixel with the largest signal is plotted against the position of the WLS fibre:
the profiles obtained in this way are similar to those in
Fig.~\ref{fig:align_profile}.  The structure of the profile closely reflects
the geometry of the PMT pixel: the peak amplitudes are proportional to
the corresponding pixel gain. Each peak is fitted with a Gaussian
function. The central position is calculated as the average of the eight 
fitted centres.

Several tests have been conducted in order to assess the
reliability of the apparatus, 
especially concerning the reproducibility of its geometry
and of the alignment procedure\cite{alfons}.  Our studies show
that the position of the PMT centre on the transverse plane is known to
within $12.5~\mu$m along the $x$ axis and to within $23.0~\mu$ along the
$y$ axis.  Moreover, no appreciable effect from any rotation around the
PMT symmetry axes has been observed.

\subsubsection{Photon Statistics}\label{sec:photon-stat}

\begin{figure}
\centerline{
\includegraphics[width=\textwidth]{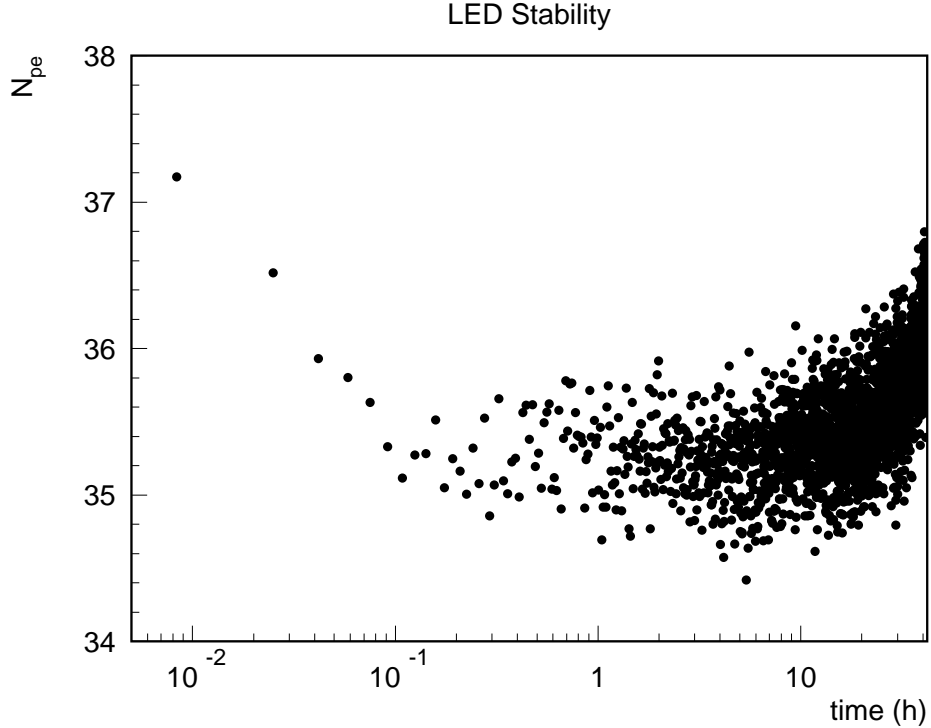}
}
\caption{This plot shows the evolution of the number of photoelectrons
$\mathrm{N_{pe}}$ on a
single PMT pixel when the fibre is not moved. Each point represents
the data of a one minute run. Once
the pulser is started its light output decreases by 5-10\% in less then
10~min. After this initial period of rapid change the output slowly varies 
by less than 1\% over several hours.}
\label{fig:led}
\end{figure}

In order to make measurements of gain and efficiency, it is necessary to
have an estimate of the amount of light impinging on the
photocathode. It is possible to make such an estimate by assuming
that the number of photoelectrons
$\mathcal{N}$ emitted at the photocathode is described by Poisson
statistics and that the amplification processes initiated by different
photoelectrons along the dynode multiplicative system are
independent. In this case the mean value ($\mu$) and the RMS ($\sigma$)
of the pedestal-subtracted charge distribution of the PMT are given by:
\begin{equation}
\mu   = \mathcal{G}\mathcal{N}
\qquad
\sigma =\mathcal{G}\sqrt{\mathcal{N}}
\end{equation}
where $\mathcal{G}$ is the PMT gain. 
The number of photoelectrons and the PMT gain are then
\begin{equation}
\label{equat:photon-stat}
\mathcal{N} = \left(\frac{\mu}{\sigma}\right)^2,
\qquad
\mathcal{G} = \frac{\mu}{\mathcal{N}}=\frac{\sigma^2}{\mu}.
\end{equation}
This method, referred to as the `photon-statistics method',
is the one used in the following paragraphs to calculate gain and number
of photoelectrons for any analysed PMT.  
The method is known to be inaccurate in several respects:
\begin{itemize}
\item
It does not take into account the fluctuations in the number of electrons
produced at the first dynodes, i.e.\ the width of the one-photoelectron peak).
Due to the sensitivity of our electronics we could not measure these
fluctuations and did not correct for them.
\item
If the tube response is non-linear, $\mathcal{G}$ becomes a function of 
$\mathcal{N}$ and Eqs.~(\ref{equat:photon-stat}) are no longer strictly valid.
However, the measurements of efficiency and gain were all done at a
level of approximately 10 p.e. where the tubes are in their linear regime.
\end{itemize}
It is still meaningful to use Eqs.~(\ref{equat:photon-stat}) 
in order to separate gain and number of photoelectrons to compare 
different PMTs.
We have verified that the LED pulser is a stable light
source (Fig. \ref{fig:led}); therefore the number of photoelectrons
gives us a measure of the relative efficiency of the pixels within a
tube.  The
efficiency here is a product of the collection efficiency (CE) and the
quantum efficiency (QE). An absolute measure of the efficiency is not
possible as the light source was not calibrated.  The absolute value of
the gain is estimated to be accurate to within approximately $20\%$.

\subsection{Results}

\subsubsection{Gain dependence on high-voltage}\label{sec:gain_vs_hv}

Scans at approximately 10 p.e.\ have been used in order to minimise possible 
effects arising from non-linearities, but still having sufficient sensitivity.
The gains for all pixels of the tested PMTs have been
measured at different high voltages (HV).

\begin{figure}
\centerline{
\includegraphics[width=0.95\textwidth]{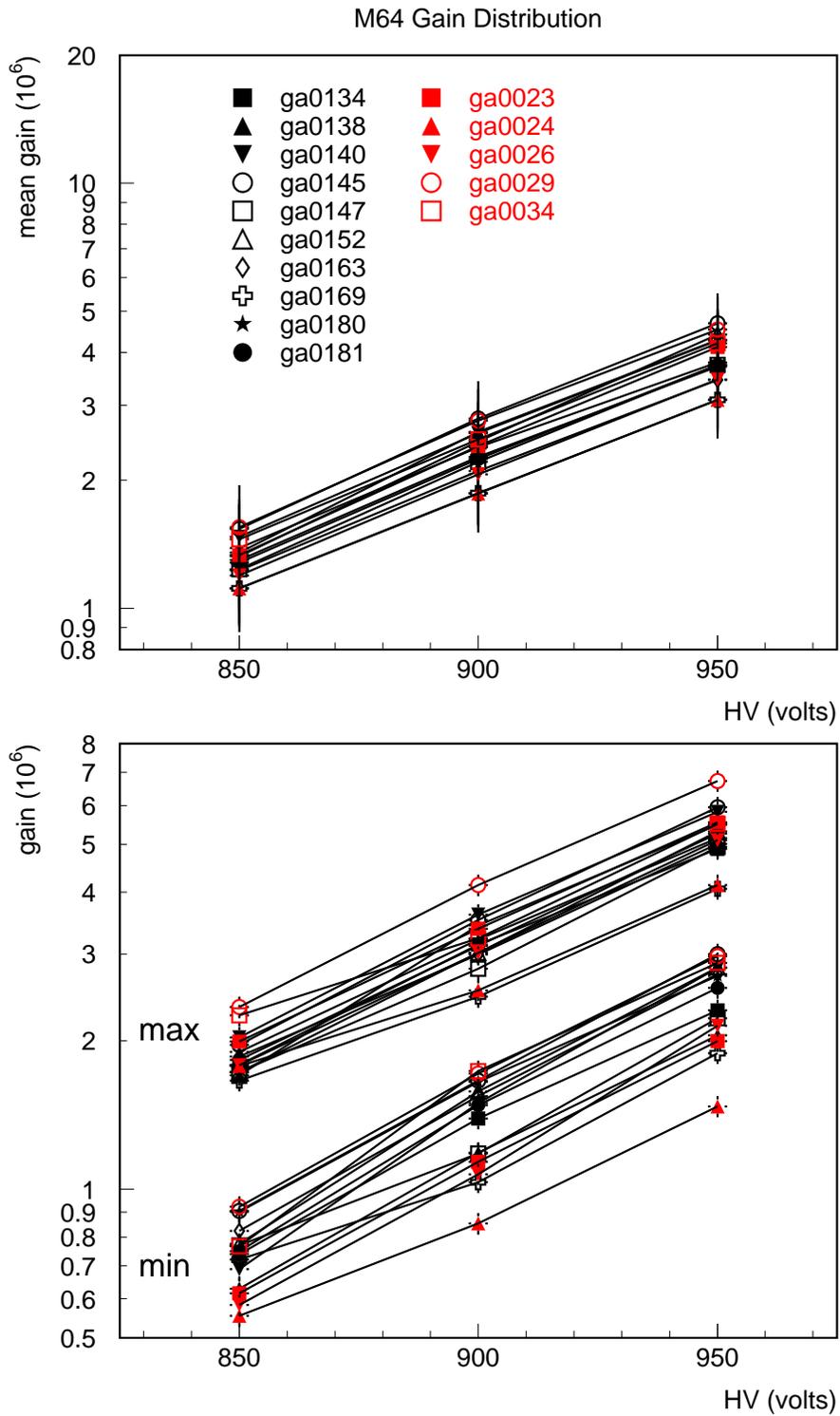}
}
\caption{The upper plot shows the average gain of all tubes as a
function of HV. 
The lower plot shows the gain of the lowest and highest gain pixel in
each of the tubes as a function of the HV. In both plots the lines
connect points belonging to the same tube.}
\label{fig:gain_vs_hv}
\end{figure}

The plots in Fig.~\ref{fig:gain_vs_hv} show the average gain of all
pixels for all of the 15 tubes as a function of HV ($850,900,950$~V).  
The gains of the
pixels with the highest and lowest gains in each tube are also shown. The
average gain increases from about $1.3\times 10^6$ at $850$~V to about
$3.9\times 10^6$ at $950$~V. Although MINOS running will require gains
of  $1.0\times 10^6$, corresponding to high voltages of lower than
850V, the sensitivity of the
electronics used in this test was not sufficient to evaluate the PMTs at
that voltage.

\begin{figure}
\centerline{
\includegraphics[width=\textwidth]{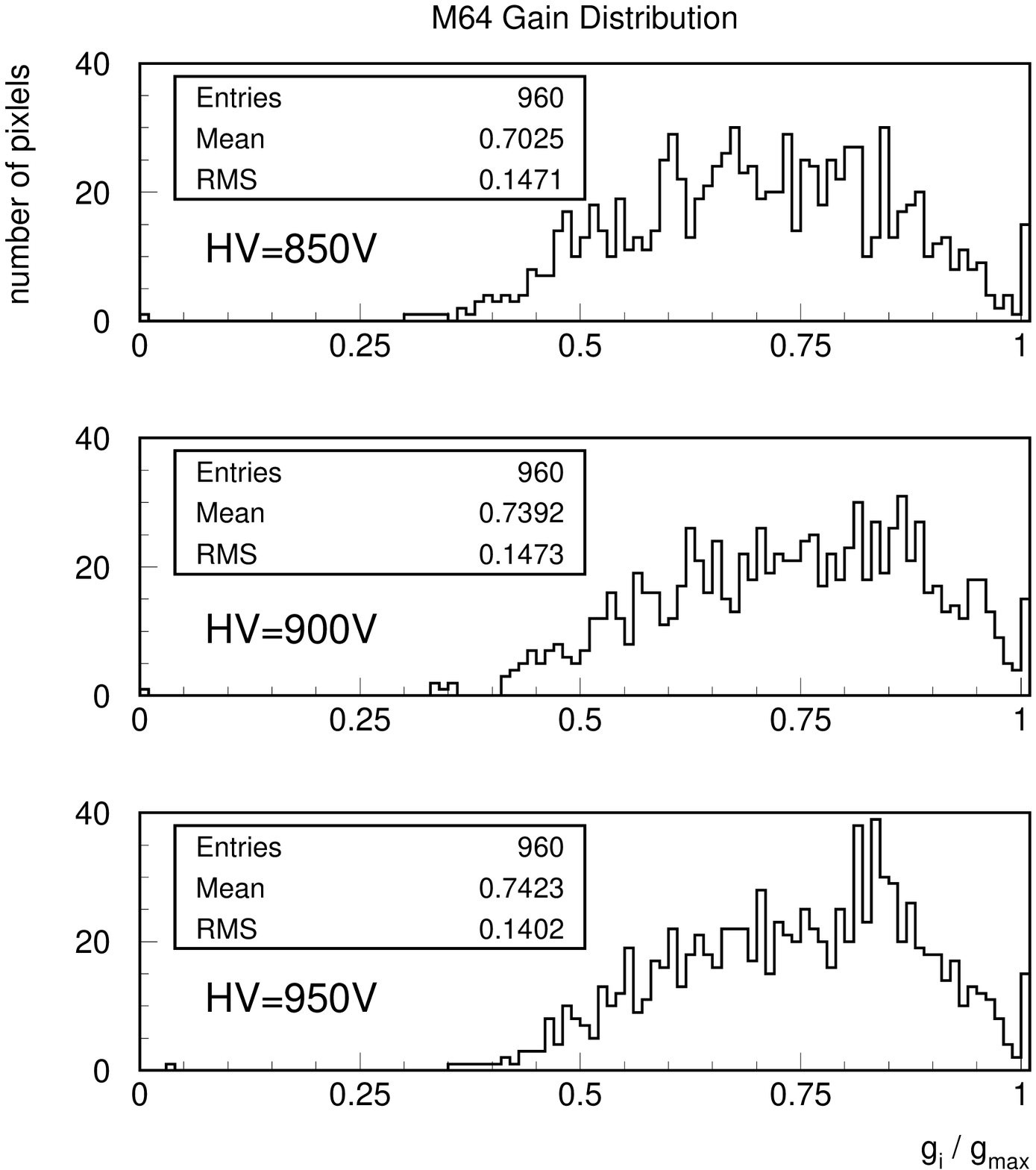}
}
\caption{Ratio between the gain of the considered channel and the
maximum gain for the PMT to which the channel belongs. The peak at one
contains the 15 entries where the highest pixel gain was divided by itself.}
\label{fig:gain_gi_over_max}
\end{figure}

The nominal gain spread 
($\mathcal{R}_{\mathcal{G}}=\mathcal{G}_{min}/\mathcal{G}_{max}$, the
gain ratio of the minimum and maximum gain pixels) for
M64 PMTs is 1:3 ($0.33$).  The average ratio measured is about 
$0.4$ at $850$~V and reaches $0.46$ at $950$~V with small differences
between different
tubes. The fact that $\mathcal{R}_{\mathcal{G}}$ decreases with
decreasing HV is probably an artifact of saturation effects,
which become more important at higher gain. Similar features can be
recognised in Fig.~\ref{fig:gain_gi_over_max}, which shows the ratio
$\mathcal{R}_i^{(k)}=\mathcal{G}_i^{(k)}/\mathcal{G}_{max}^{(k)}$
between each channel gain and the maximum gain of the same PMT, where
$i$ runs over the PMT index and $k$ runs over the pixel number. The 15
entries at $\mathcal{R}_i^{(k)}=1$ in each histogram correspond to the
15 maximum gain pixels.

\subsubsection{Cross-talk}

\begin{figure}
\centerline{
\includegraphics[width=\textwidth]{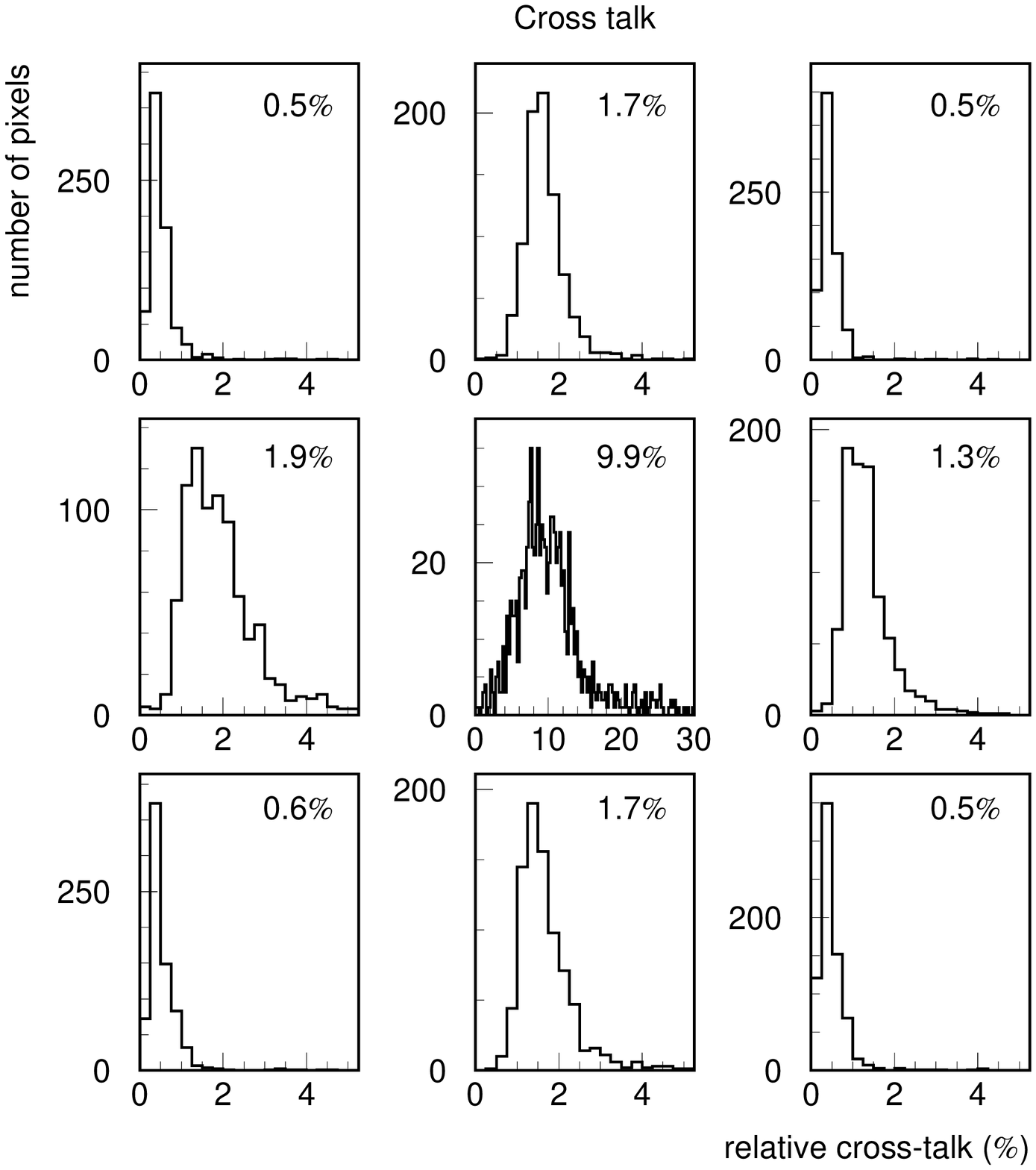}
}
\caption{Cross-talk $\mathcal{C}$ (in percent) into the neighbouring
pixels. The entries in the histograms are the means of the
corresponding distributions. 
The central histogram shows the distribution of the total cross-talk for
all of the pixels measured. It shows that on average $\sim10\%$ of the
total signal is seen in pixels that are not illuminated. However for
some pixels this increases to more than 25\%. Any of the border histograms
in the $3\times 3$ matrix shown quantifies the cross-talk into a
specific pixel. For example: the left middle histogram gives the
distribution of the cross-talk signal induced by the illuminated pixel
in the pixel at its left (if it exists); similarly, the plot at the
upper-right corner gives the distribution of the cross-talk induced in the
upper-right neighbour pixel; and so on.  }
\label{fig:xtalk_neighbours}
\end{figure}

As all channels were read out constantly during the PMT scans, it was
straightforward to measure the PMT cross-talk. The cross-talk
$\mathcal{C}$ for a fibre position $j$ has been defined to be
\begin{equation}
\label{eq:xtalk}
\mathcal{C} = \frac{1}{Q_j}\sum\limits_{i\neq j }Q_i,
\end{equation}
where $Q_j$ is the charge of channel $j$ and the sum runs over a certain
class of neighbouring channels. All charges are averaged over 10,000
events.  The cross-talk into direct or diagonal channels as well as the
total cross-talk has been measured.  Most of the cross-talk is into
direct neighbours and only much reduced cross-talk into diagonal
pixels was observed.  
Results are summarised in Fig.~\ref{fig:xtalk_neighbours},
where the cross-talk for all the tested phototubes is displayed. The
total cross-talk is about $10\%$ on average, with long tails extending to
$\sim 25\%$.  The cross-talk into non-neighbouring pixels sums up to about
$3\%$.

\begin{figure}
\centerline{
\includegraphics[width=\textwidth]{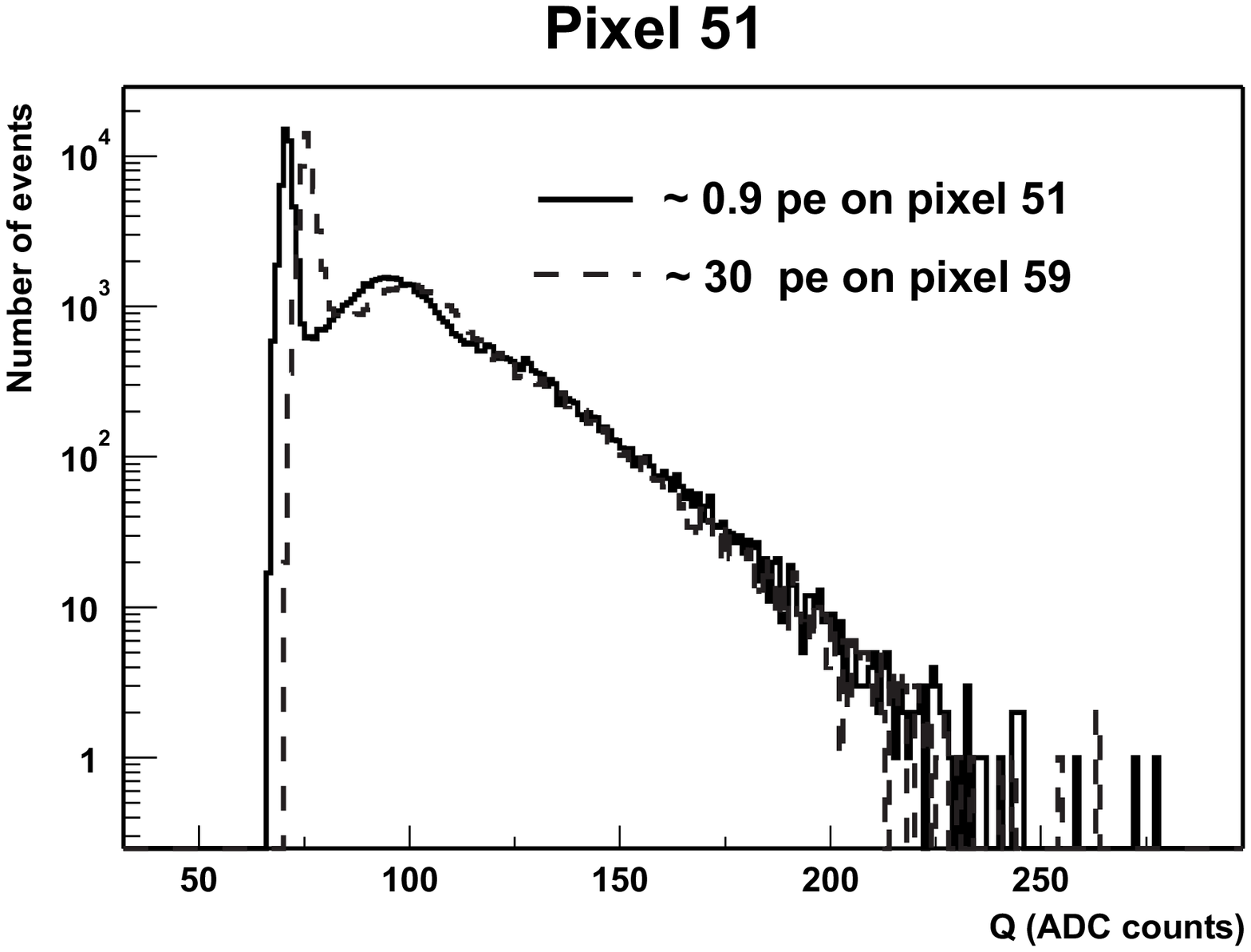}
}
\caption{Charge distributions for pixel 51. 
Two spectra are plotted.
The solid spectrum shows the distribution when the pixel itself is
pulsed with an average of 0.9 photoelectrons. The left peak is the
pedestal which has not been subtracted. The single photoelectron peak is
visible at 100 ADC counts.
The dashed distribution
has been recorded when the neighbouring pixel 59 was illuminated with
approximately 30 photoelectrons. The distribution has a similar single
photoelectron peak. The pedestal however is shifted,
indicating that additional charge was measured in pixel 51, even without
a photon present.}
\label{fig:xtalk_spectra} 
\end{figure}

As an example we have plotted the charge distributions for pixel 51 
in Fig.~\ref{fig:xtalk_spectra}. This measurement was performed with
a modified base (voltage divider ratio $3:2:2:1:1:1:1:1:1:1:1:1:1$) and
using a high bandwidth $\times 10$ amplifier to achieve the required
sensitivity. Two spectra are compared:
The solid spectrum shows the distribution when the pixel itself is
pulsed with an average of 0.9 photoelectrons. The dashed distribution
had been recorded when the neighbouring pixel 59 was illuminated with
approximately 30 photoelectrons. From plots like this one can determine
the kind of cross-talk present:
\begin{enumerate}
\item
A distinct single photoelectron peak is visible when illuminating a
neighbouring pixel. 
This suggests that some of the photons coming through the WLS
fibre are not producing a photoelectron at the fibre location, but on
adjacent pixels. This can be explained by the divergence of the light
leaving the fibre and internal reflections in the PMT glass window.
\item
The pedestal peak visible in Fig.~\ref{fig:xtalk_spectra} is 
different from the one observed when pulsing the pixel directly.
It is shifted to higher values. This shift is proportional to the light level.
It is an indication that a constant fraction of charge is seen at the
neighbouring anode due, presumably, to charge leaking across to the
neighbouring dynode structure during multiplication.

\end{enumerate}

\subsubsection{Linearity studies}
\label{sec:non-lin}

Photomultiplier tubes are in general reasonably linear devices: the
charge at the anodes is proportional to the amount of light incident on
the cathode. Nevertheless,
when the incident light is too intense or when the bias voltage is too
high, significant deviations from the ideal linear response are
observed. This is mainly related to the dynode chain linearity characteristics,
which are independent of the incident light wavelength and which, at
fixed high-voltage values, are functions only of the peak current
passing through the last dynode stages\cite{Phillips-book,hamamatsu-book}.
Space charge effects are very important for PMTs
operated in pulsed mode when intense light pulses can reach the
photocathode and induce large current flows through the last dynode
stages in addition to leaving charge between these dynodes.  
The magnitude of the consequent saturation effect is related to
the internal tube geometry, the design of the resistive bleeder
circuit, as well as to the electric field intensity and its
spatial distribution.

%The PMT linearity is limited by the voltage divider and space charge
%effects\cite{Phillips-book,hamamatsu-book}.
%In DC mode linearity is mainly affected by the design of the
%bleeder circuit. 

\begin{figure}
\centerline{
\includegraphics[width=\textwidth]{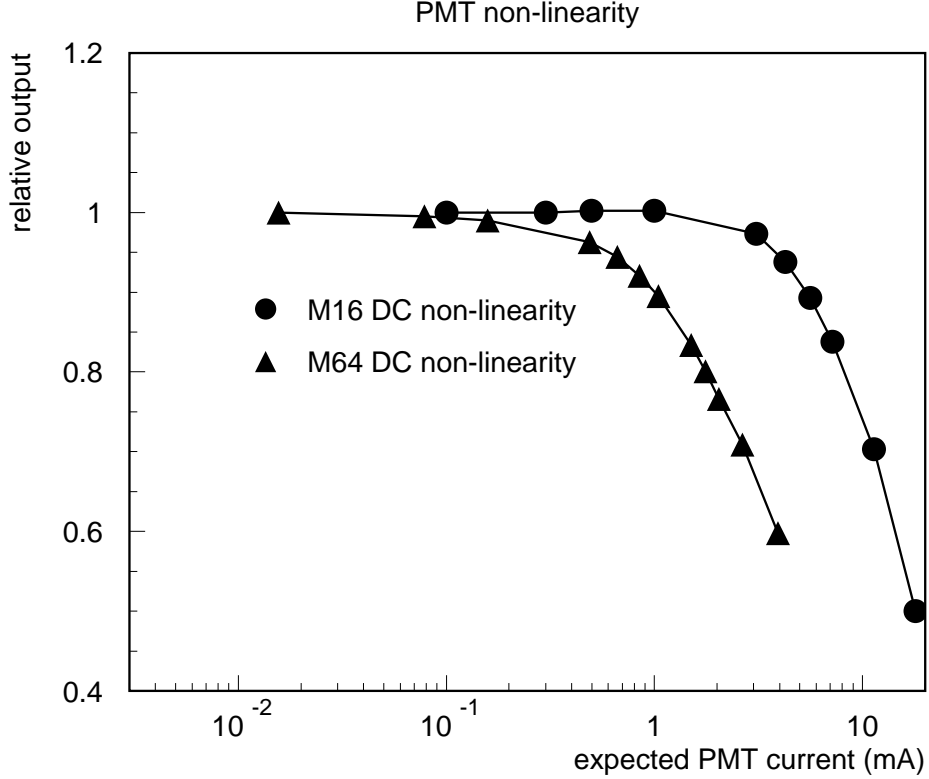}
}
\caption{Relative outputs for M16 and M64 phototubes as a function
of the instantaneous PMT current.}
\label{fig:m64_m16_nonlin}
\end{figure}

The use of a tapered bleeder circuit (see Sec.~\ref{sec:m64})
reduces such effects, as the large potential differences in the last few
stages of the amplification chain enhance the voltage gradient
where the electron space density becomes higher.  Nonetheless, the M64
phototubes still exhibit a considerable non-linear response, especially
when compared to M16 PMTs which is a consequence of the M64's
comparitively smaller pixel size. 
The different behaviour of the two types of
phototubes is very clear in Fig.~\ref{fig:m64_m16_nonlin}, where the
relative outputs of the two photodetectors are shown as a function of the
instantaneous PMT current when one pixel is illuminated (Hamamatsu
data): while M16s are essentially linear for currents up to about
$1$~mA, M64s depart from linearity already at about $0.1$~mA.

\begin{figure}
\centerline{
\includegraphics[width=\textwidth]{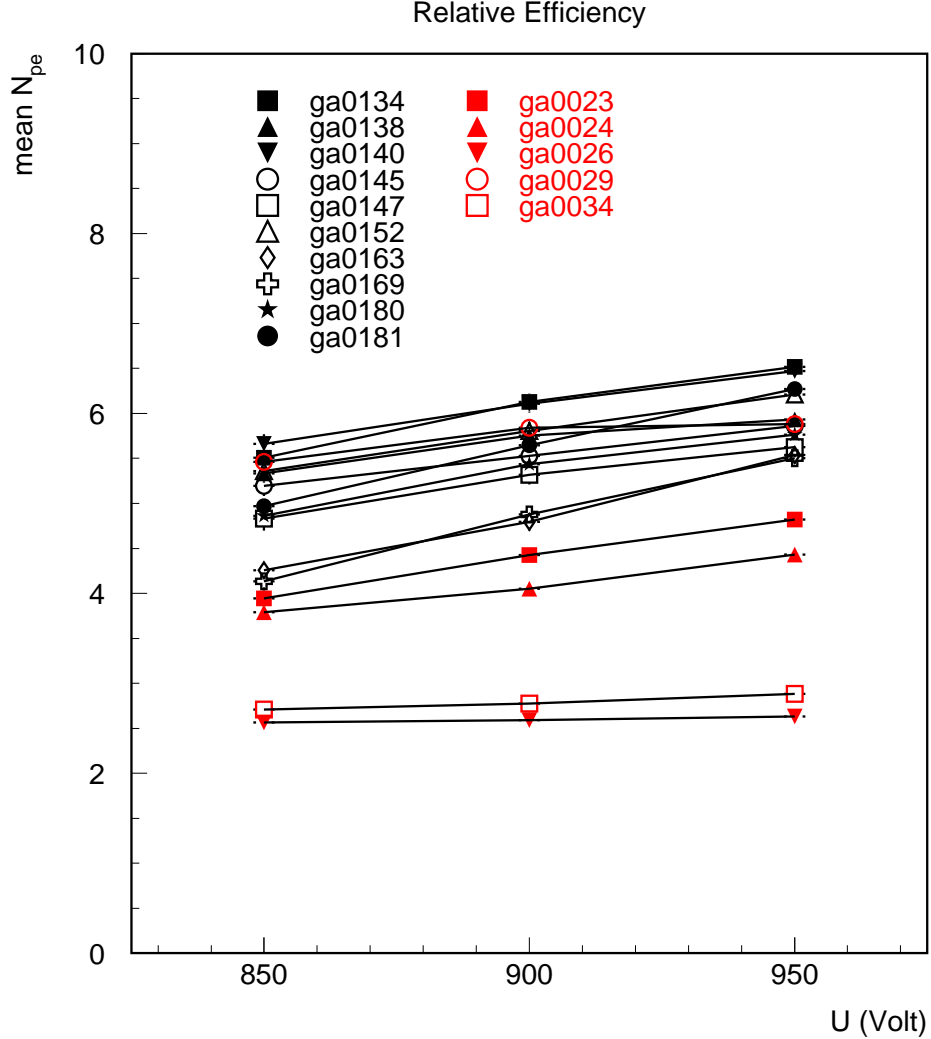}
}
\caption{The average number of photoelectrons $N_{pe}$ (as calculated from
the photon-statistics method) versus the PMT high voltage for the $15$ M64s
under test.}
\label{fig:ngamma_vs_hv}
\end{figure}

\begin{figure}
\centerline{
\includegraphics[width=\textwidth]{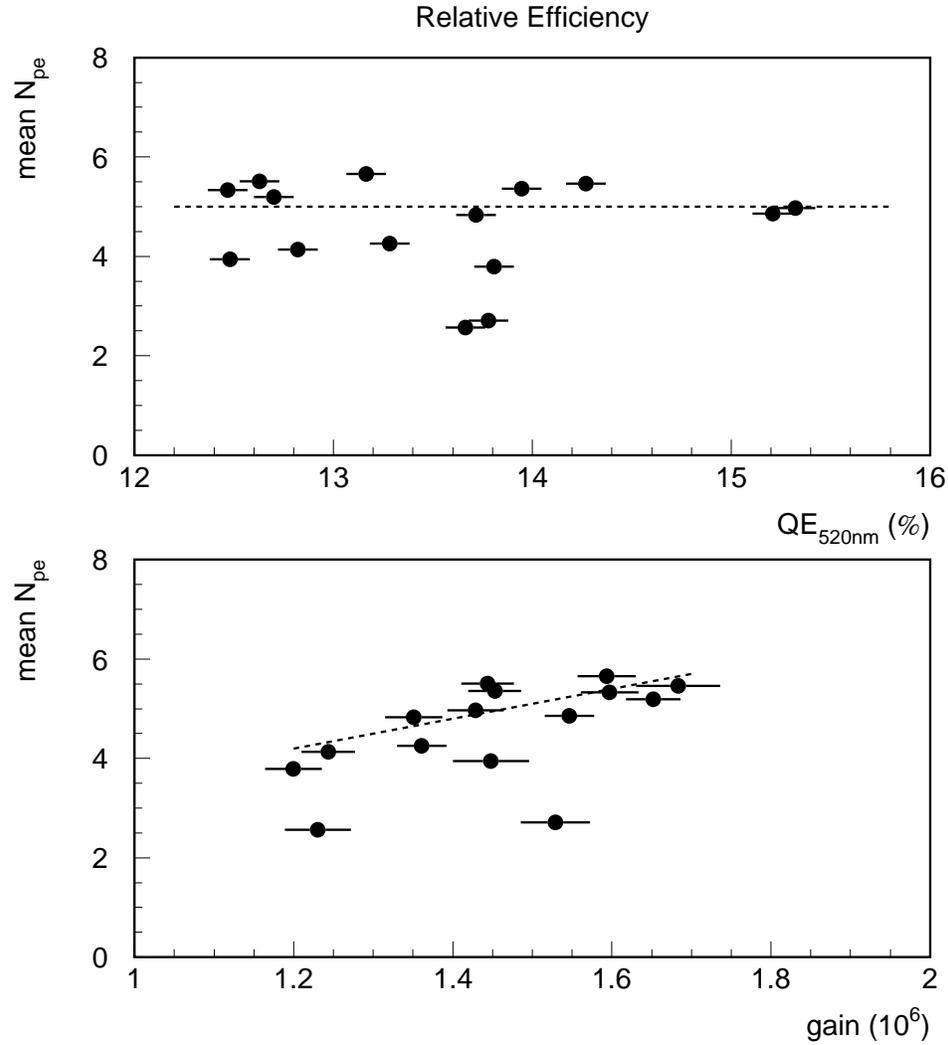}
}
\caption{Calculated number of photoelectrons as a function of QE at
$520$~nm (top) and of the computed PMT gain (bottom); QE$(520$~nm$)$
comes from Hamamatsu measurements.} 
\label{fig:ngamma_qe_ce}
\end{figure}

\begin{figure}
\centerline{
\includegraphics[width=\textwidth]{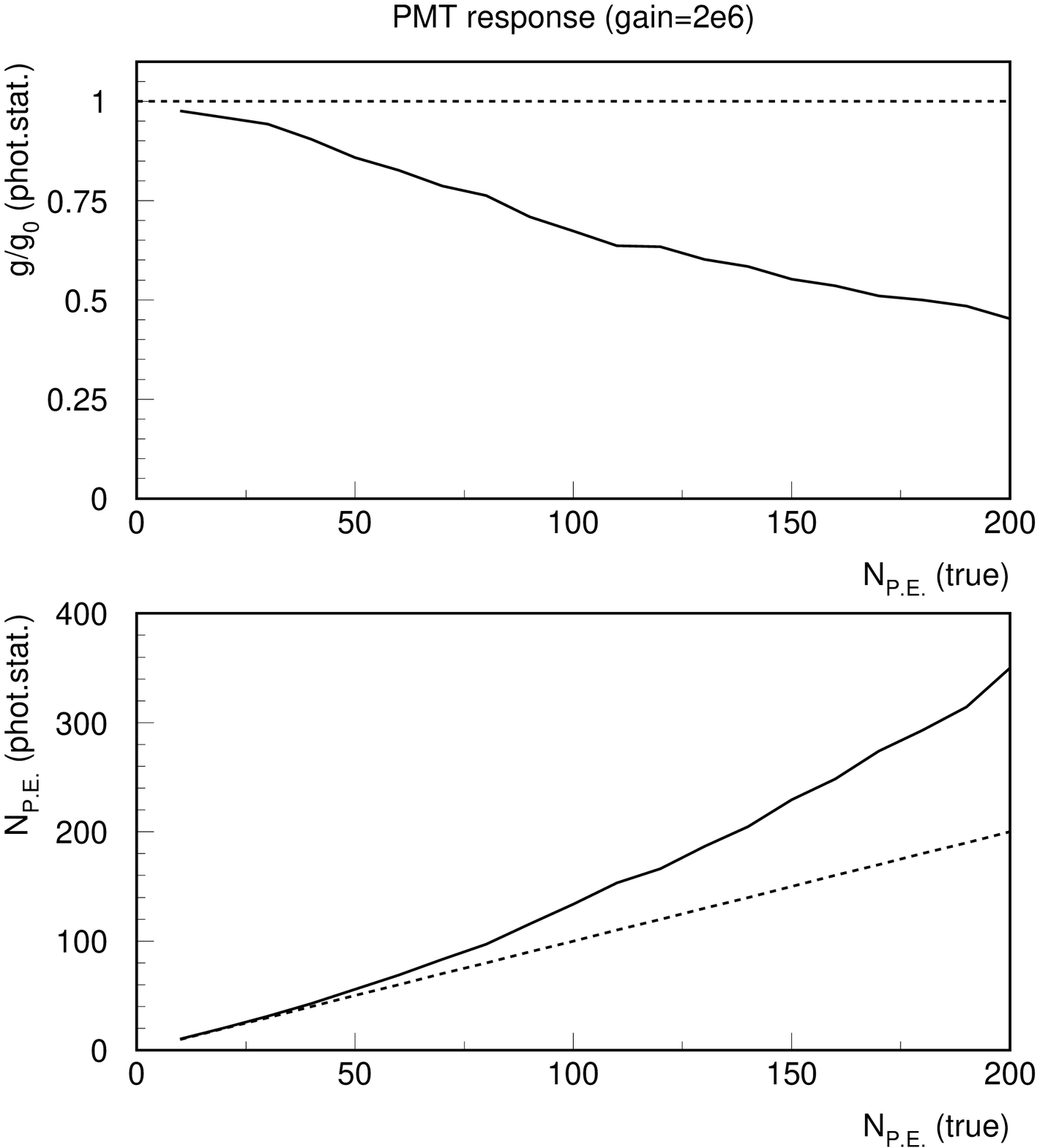}
}
\caption{These plots show the number of photoelectrons and the gain as 
calculated by photon statistics as a function of the real number of 
photoelectrons.
They are a result of a simulation which includes realistic pulse shapes for
the incident light and the current non-linearity as measured by Hamamatsu.
The measured charge distribution is squeezed together due to the
non-linearity,
resulting in a higher calculated number of photo-electrons and a lower gain.}
\label{fig:non-linearity-sim}
\end{figure}

The photon-statistics method, described in Sec.~\ref{sec:photon-stat},
relies on the assumption that the PMT is a linear device. Assuming that
the light source is stable, any deviation from linearity should have a
direct effect on the quantities calculated in
Eq.~(\ref{equat:photon-stat}). This is what is observed in
Fig.~\ref{fig:ngamma_vs_hv}, where $\mathrm{N_{pe}}$, the number of
photoelectrons, is plotted as a function of HV. The spread in
$\mathrm{N_{pe}}$ at fixed HV reflects a general change in light level,
when testing different sets of tested tubes. (We had to demount and remount
the fibres in the attenuation box leading to a change in overall light
level.) 
It appears
that the calculated number of photoelectrons is a function of the applied
voltage; but of course the light level cannot really depend on the voltage.
To verify that this dependence is an artifact of the photon-statistics
computation and not an effect correlated with the PMT quantum efficiency
and collection efficiency, in Fig.~\ref{fig:ngamma_qe_ce} $\mathrm{N_{pe}}$ is
plotted as a function of QE at $520$~nm and of the calculated gain
$\mathcal{G}$ separately: $\mathrm{N_{pe}}$ 
seems to be independent of QE, while
a clear dependence on $\mathcal{G}$ is observed.  Simulations have
confirmed that tube non-linearities can influence the calculations of
$\mathrm{N_{pe}}$ in the observed way when the light level is 30-50
photoelectrons as shown in Fig.~\ref{fig:non-linearity-sim} 

At light levels as low as 3-5 photoelectrons the PMT
gain fluctuations (width of the single-photoelectron peak) become dominant
and influence the calculation. This width varies with gain.
The relative width of the single-photoelectron peak becomes narrower as
the gain increases, leading to a higher calculated number of
photoelectrons. 

\subsubsection{Sensitivity to Magnetic Fields}

To test the potential consequences of a stray magnetic field on the PMT
performances, a Helmholtz coil was used to generate a homogeneous
field of up to $40$~gauss\cite{alfons}.  A full scan
(a total of $40\times 40$ points, $5\times 5$ points per pixel) of several
tubes was performed for each value of the magnetic field.  No
effect was observed for a field orientation perpendicular to the
PMT longitudinal axis ($z$), either on the tube gain or on collection
efficiency. However, if the field is parallel to the $z$-axis,
significant effects on the collection efficiency of the outer pixels are
observed at intensities of $7$~gauss and above.
Cross-talk was observed not to be affected by the presence of a magnetic
field, regardless of its orientation.

%% file: smry.tex
\section{Conclusions}\label{sec:concl}
We have measured the properties of the Hamamatsu R5900-00-M64
multi-anode photomultiplier tube and find that these tubes are
well suited for the MINOS experiment. We have found that the main
properties of the tube are:
\begin{itemize}
\item[-]
The quantum efficiency at 520 nm is $13\pm0.5 \%$
\item[-]
The average tube gain is 1.3 $\times 10^6$ at 850 V while the ratio of
minimum to maximum pixel gain is better than 1:3.
\item[-]
There are only small variations in the tube efficiency.
The typical variation has been measured to be $20\%$. 
\item[-] 
There are already sizable non-linearities at a tube gain of $10^6$.
\item[-]
When placing a 1.2 mm acrylic fibre on the photocathode, the cross-talk
to neighbouring pixels is on average $10\%$ which includes
contributions from both optical and charge cross-talk
\item[-] 
Typical dark count rates at $20^\circ\mbox{C}$ are less than 10 Hz per
pixel for
signals above 1/3 of a photoelectron.
\item[-]
The tubes are to a large extent insensitive to magnetic fields of up
to 5~gauss. 
\end{itemize}
\begin{ack}
We would like to thank Hamamatsu for their cooperation and for
providing extensive test data for their PMTs. We also thank all the
technicians and engineers who helped in preparing these measurements. 
This work was supported in part by PPARC, the UK Particle Physics
and Astronomy Research Council, and by DOE, the United States Department
of Energy.
\end{ack}